# The boundary and continual transfer phenomena in fluids and flows

S.L. Arsenjev, I.B. Lozovitski[1], Y.P. Sirik

*Physical-Technical Group*
*Dobroljubova street 2, 29, Pavlograd, Dnepropetrovsk region, 51400 Ukraine*

The clearing up of a wave nature of the energy and mass transfer phenomena in classical expressions of the molecular-kinetic theory has allowed to find a quantitative measure of intensity of processes of a thermal conductivity, viscosity and diffusion in conditions of a thermally nonequilibrium and heterogeneous composition continuum. It is rotined, that the appearance of a temperature drop in fluid stipulates the appearance of the continuum stratification and formation of the flowing "bodies" interacting among themselves. It is rotined, that the known expressions for a thermal conductivity, viscosity, diffusion and heat convection had been obtained for a thermally equilibrium and homogeneous continuum and produce a maximum quantity of intensity of the transfer processes. The introduced expression is usable for a quantitative estimation of intensity of the transfer processes in fluid and its streams in conditions of the non-stationary heat exchange in natural conditions and technical problems.

**Introduction**

The kinetic theory of gases is the most developed and oldest discipline of the molecular-kinetic theory of matter. To middle of the XX century, the basic laws of the energy and mass transfer phenomena in the gas medium were ascertained within the framework of this theory. For ascertainment of these laws, it was accepted, that the gas medium is homogeneous, continuous, is in stationary state of dynamic balance, appropriate for Maxwell's function of the distribution of molecules by velocities of its motion. The detected experimentally dependence of the transfer phenomena on temperature was taken into account by Suzerland's formula defining dependence of a free length of the gas atoms (molecules) on its temperature [1]. At the same time, it was recognized that «…the theory though and correctly reflects a progress of processes but one is not realistic in the numerical data» [2]. The results of the careful experiments of the gas stream moving with the heat exchange in the pipe [3, 4] testifies that the greater is the temperature difference between the pipe wall and flow, the less is quantity of the heat transfer and hydraulic friction coefficients. About the offered empirical methods, taking into account of the abovementioned appearances, the authors [5] testify themselves that its applying «can reduce to incorrect calculation.» If to this to add such problems: why the flame tongue is localized in an atmosphere, why the atmospheric and ocean currents are localized in the jet streams and ones are moved at huge distances and all that, so the unsatisfactory state of the question of quantitative estimation of intensity of the transfer phenomena in fluid will be detected in full measure.

---

[1] Phone: (38 05632) 38892, 40596
E-mail: loz@inbox.ru

**Approach**

Let's consider a following situation: two gas mediums, reposing in a zero gravity of identical composition under identical pressure but having different temperature, are entered contacts. Let's call one of these mediums "hot" and the other "cold". The distribution of molecules on its motion velocities in each of the viewed mediums matches to Maxwellian function. According to this law, the hot gas medium contains a few molecules having velocity equal to the average velocity of molecules of the cold gas medium. And vice versa, the cold gas medium contains a few molecules having velocity equal to the average velocity of molecules of the hot gas medium. But the average velocity of molecules of the hot gas medium is much more of the average velocity of the cold gas medium. At the moderate value of pressure and temperature, the number of the collisions of one molecule with the adjoined molecules for both mediums is equally though density of hot medium is much less than density of cold medium. The last combination implies, that on boundary of contact of the considered mediums each molecule of hot medium sequentially collides with several molecules of cold medium. Such boundary interaction essentially differs from interaction of molecules inside each of the considered mediums. Maxwellian distribution, as the most probable, means a uniform distribution of molecules on its velocities in space of this or that gas medium. Thus the motion of molecules in medium has the penetrating (diffusion) nature. On the other hand, the correspondence of the internal energy (oscillation) of molecules to the kinetic energy of its translational motion is property of dynamic self-stabilization of the initial state of each of the considered mediums. The last property is similarly to the first mechanics principle in certain degree and one can be expressed so: every molecular-kinetic medium maintains the equilibrium according Maxwell's state in absence of external actions.

In contrast to the penetrating interaction of molecules inside each of the contacting gas mediums, the contact interaction of these mediums has a reflective nature to the certain extent. At the same time intensity of the transfer processes of the thermal and mechanical energy and also of the mass (diffusion) is descended depending on increasing of the temperature drop between contacting mediums.

Taking into account on elastic character of interaction of molecules with each others both inside of the gas medium and on boundary with other medium, it is expedient to suppose that the transfer phenomena of energy and mass through the boundary of the gas mediums have the wave nature. From this position, we shall consider expressions, known in the molecular-kinetic theory for:

heat conductivity $\quad\quad\quad\quad \lambda = \frac{1}{3} \cdot l \cdot \bar{c} \cdot \rho \cdot c_v$,

viscosity $\quad\quad\quad\quad\quad\quad \mu = \frac{1}{3} \cdot l \cdot \bar{c} \cdot \rho$,

diffusion $\quad\quad\quad\quad\quad\quad D = \frac{1}{3} \cdot l \cdot \bar{c}$.

If both parts of expression for diffusion to multiply on $\rho$, then all three these expressions will have identical dimensionality of specific stream (fluence) of energy. At the same time, all three these expressions will contain a product $\bar{c} \cdot \rho$, which one characterizes the wave impedance of the medium. That is, the wave nature of the transfer phenomena in the gas medium is embodied in the formulas of the molecular-kinetic theory, despite of its appreciably statistical nature.

**Solution**

According to the wave approach [6] on the boundary surface of two mediums having the different wave impedance, the waves of energy out of medium with the high wave impedance are reflected partially or totally from the contact boundary of mediums, and fractionally transits into medium with the smaller wave impedance. Thus the condition of the continuity of mediums on boundary of its partition is remained.

The expression for the reflection coefficient of the energy flow on a boundary of mediums with the different wave impedance has well-known view:

$$K_{br} = \left| \frac{z_1 - z_2}{z_1 + z_2} \right|^2 ,$$

where $z = \rho \cdot c$ - wave impedance of the medium, $c$ – sound velocity.

Accordingly expression for the conductivity coefficient of boundary of mediums with the different wave impedance:

$$K_{bc} = 1 - K_{br} = 1 - \left| \frac{z_1 - z_2}{z_1 + z_2} \right|^2 .$$

As applied to the gas mediums the conductivity coefficient represents the quantitative performance of degree of depression (diminution) of intensity of the transfer phenomena in molecular-kinetic medium.

So, for the contacting gas mediums of different composition under the different temperature and the same pressure, it is rightly the expression:

$$K_{bc} = 1 - \left| \frac{\rho_1 \sqrt{k_1 \cdot g \cdot R_1 \cdot T_1} - \rho_2 \sqrt{k_2 \cdot g \cdot R_2 \cdot T_2}}{\rho_1 \sqrt{k_1 \cdot g \cdot R_1 \cdot T_1} + \rho_2 \sqrt{k_2 \cdot g \cdot R_2 \cdot T_2}} \right|^2 , \quad (1)$$

where $\rho_1$, $\rho_2$ – weight density of mediums, $k_1$, $k_2$ – adiabatic exponent of mediums, $R_1$, $R_2$ – specific gas constant of mediums, $T_1$, $T_2$ – temperature of the contacting mediums.

The expression (1) was applied for the quantitative estimation of the relative heat exchange between the gas stream and the pipe wall both at heating and at cooling of stream in the form:

$$K_{bc} = Nu_{ex}/Nu_c , \quad (2)$$

where $Nu_{ex}$, $Nu_c$ – experimental and calculation quantity of the Nusselt's number accordantly.

Comparison of results of calculation with results of known experiments [3, 4] testify that the expression (1) correctly and precisely enough plays back the results of experiments. Thus the expression (1) was utilised in two simplified modifications:
- at heating of stream by the pipe wall

$$K_{bc} = 1 - \left| \left(1 - (T_f/T_w)^{0.5}\right) / \left(1 + (T_f/T_w)^{0.5}\right) \right|^2 , \quad (3)$$

- at cooling of stream by the pipe wall

$$K_{bc} = 1 - \left|\left(1 - (T_w/T_f)^{0.5}\right)/\left(1 + (T_w/T_f)^{0.5}\right)\right|^2 , \qquad (4)$$

where $T_f$, $T_w$ – the temperature is average on section of stream and the temperature of the pipe wall accordingly.

It is necessary to note that for the equilibrium conditions, when the temperature drop in medium is absent, the boundary conductivity coefficient $K_{bc}$ becomes equal to unity. It match to experimental data [3, 4] and testify that the Nusselt's number usually used is deduced for the equilibrium conditions when the heat exchange is absent.

Similarly to this and the classic expressions of the molecular-kinetic theory for transfer of thermal and mechanical energy and the mass (diffusion) are deduced for the equilibrium conditions when there is absent the temperature drop in the considered medium. Generally when there is the temperature drop in molecular-kinetic medium, there are boundaries in it on which one the intensity of processes of transfer is descended. For these conditions, the classical expressions become:

$$\lambda_b = K_{bc} \cdot \lambda , \qquad (5)$$

$$\mu_b = K_{bc} \cdot \mu , \qquad (6)$$

$$D_b = K_{bc} \cdot D , \qquad (7)$$

Thus, the intensity of all processes of transfer is descended in the equal number of times under action of the temperature drop. In this connection it is necessary to distinguish conductivity in the thermally equilibrium molecular-kinetic continuum at given temperature that can be estimated by means of classical expressions, and alongside with it, it is necessary to distinguish the boundary conductivity when in indicated medium, the boundaries are arisen under action of the temperature drop, dividing the continuum on the fluid "bodies". The intensity of the boundary transfer is determined by expressions (5, 6, 7).

The uniformity of change of intensity of the transfer processes under the thermally nonequilibrium conditions returns us to the consideration of the gas stream motion in pipe. This is interlinked with that at the heat exchange (of stream with the pipe wall), the hydraulic friction coefficient is necessary to multiply on the conductivity coefficient just as the heat exchange coefficient. In this case only, results of well-known experiments [5], in which one the lowering of the heat exchange intensity as contrasted to result of calculation was accompanied simultaneously by the lowering of the hydraulic friction coefficient, will be clear.

**Discussion of results**

The wave character of the contact interaction of solid bodies and distribution in its of mechanical energy is well learnt at present time [7]. The mechanics of the contact interaction implies presence of bodies entering the contact interaction among themselves. As against it, the fluid possessing ability to conduct the mechanical energy in the form of elastic waves as well as the solid body, itself is capable to be structured on the bodies interacting among themselves. This property of fluid is consequence that it has the greater number of degrees of freedom as contrasted to solid body. The reasons of the self-structurization of a gas and liquid continuum into the flowing bodies are spatial heterogeneity of composition and the thermal non-equilibrium of continuum. In oceanology and meteorology, for example, these reasons explain an existence of the underwater and atmospheric sound channels owing to the boundary refraction. The wave approach offered in this work allows to ascertain not only this phenomenon but also a reason of existence of the currents in hydrosphere and atmosphere. The sharp decrease of intensity of the transfer processes of energy and matter on boundary of heterogeneous composition and thermal non-equilibrium leads to the stratification of the water and air continuum. The flowing bodies nascent in the result of structuring of continuum are propelled in the form of the jet flows under action of the gravitation forces and the pressure drops. Owing to a sharp decrease of intensity of the boundary processes of the energy and mass transfer, the warm and cold currents in oceans and atmosphere have global time-space nature. It is observed also the light refraction on these boundaries.

In multiplicity of the technical realizations, it is possible to cite an example illustrating the operational efficiency of the two-stroke internal combustion engine. The distinction of this type engine from the four-cycle engine implies that the processes of the exhaust of the combustion products and filling of the combustion chamber by the fresh air-fuel mixture are combined in the united blowing process. The fresh air-fuel mixture entering the combustion chamber under an overpressure should expel the combustion products and should not mix up with its. The experience displays that the intermixing of these gas mediums practically does not happen. The fresh air-fuel mixture expels the combustion products similarly to the piston out of the combustion chamber.

Strange as it may seem, the basic reason of this phenomenon is only that the temperature of the combustion products is much higher than temperature of the combustion chamber walls, and also of the fresh air-fuel mixture. At first, this sole cause practically prevents intermixing of the concerned gas mediums. Secondly, It appreciably depreciates a viscosity of the combustion products on the contact boundary with the air-fuel mixture and walls of the combustion chamber and simultaneously fortifies an integrity of medium of the combustion products, owing to the high continual viscosity. The combination of the listed features of the blowing process stipulates the slip out of the combustion products out of the combustion chamber under action of pressure of the arriving air-fuel mixture to the cylinder without essential intermixing of these gas mediums.

The phenomenon of the continuum stratification of fluid (gas, liquid) comes into being as the consequence of the thermal non-equilibrium is general at the natural and technical processes.

**Final remarks**

This work is executed initiatively and independently by the scientists of Physical-Technical Group within the framework of development of subject «Physics of motion of liquid and gas». By this work authors anticipate statement "Fluid Motion Wave Theory" in subsequent articles.

---